\documentclass[pra,twocolumn,superscriptaddress,aps,floatfix,longbibliography]{revtex4-2}

\usepackage{hyperref}
\usepackage{amsmath}
\usepackage{bm}
\usepackage{graphicx}
\usepackage[ansinew]{inputenc}
\usepackage{array}
\usepackage{color}
\usepackage{amsxtra} 
\usepackage{amstext}
\usepackage{amssymb}
\usepackage{latexsym}
\usepackage{dsfont}
\usepackage{braket}
\usepackage[caption=false]{subfig}
\usepackage[makeroom]{cancel}
\usepackage{lineno}
\usepackage{physics}
\usepackage{orcidlink}

\usepackage{dcolumn}
\usepackage{bm}
\usepackage{bbm}
\usepackage{braket}
\usepackage{mathrsfs}
\usepackage{amstext}
\usepackage{euscript}

\usepackage{hyperref}

\usepackage{txfonts}

\usepackage{xcolor}
\definecolor{kjcol}{rgb}{0.8,0,0.5}
\definecolor{pink}{rgb}{1,0,0.9}
\definecolor{dgreen}{rgb}{0,0.6,0}
\definecolor{dmag}{rgb}{0.6,0.0,0.6}

\usepackage[normalem]{ulem}

\newcommand{\ms}[1]{\mbox{\scriptsize #1}}

\newcommand{\cref}[1]{Ref.\,\cite{#1}}

\begin{document}

\title{NV-ensemble enabled microwave/NV parametric amplifier with optimal driving}

\author{Roman Ovsiannikov\orcidlink{0009-0003-9558-4132}}
\email{roman.ovsiannikov@kipt.kharkov.ua}
\affiliation{Akhiezer Institute for Theoretical Physics, NSC KIPT, Akademichna 1, 61108 Kharkiv, Ukraine}

\author{Kurt Jacobs\orcidlink{0000-0003-0828-6421}}
\email{dr.kurt.jacobs@gmail.com}
\affiliation{Advanced Photonic and Electronic Sciences Division, U.S. Army DEVCOM Army Research Laboratory, Adelphi, Maryland 20783, USA} 
\affiliation{Department of Physics, University of Massachusetts at Boston, Boston, Massachusetts 02125, USA}

\author{Andrii G. Sotnikov\orcidlink{0000-0002-3632-4790}}
\email{a\_sotnikov@kipt.kharkov.ua}
\affiliation{Akhiezer Institute for Theoretical Physics, NSC KIPT, Akademichna 1, 61108 Kharkiv, Ukraine}
\affiliation{Education and Research Institute ``School of Physics and Technology'', Karazin Kharkiv National University, Svobody Square 4, 61022 Kharkiv, Ukraine}

\author{Matthew E. Trusheim} 
\email{mtrush@mit.edu}
\affiliation{Advanced Photonic and Electronic Sciences Division, U.S. Army DEVCOM Army Research Laboratory, Adelphi, Maryland 20783, USA}  
\affiliation{Department of Electrical Engineering and Computer Science, Massachusetts Institute of Technology, 77 Massachusetts Avenue, Cambridge, Massachusetts 02139, USA}

\author{Denys I. Bondar\orcidlink{0000-0002-3626-4804}}\email{dbondar@tulane.edu}
\affiliation{Department of Physics and Engineering Physics, Tulane University,  New Orleans, Louisiana 70118, USA}

\date{\today}
\begin{abstract} 

In our recent study~[\href{https://doi.org/10.48550/arXiv.2601.03407}{arXiv:2601.03407}] we showed that a hybrid non-degenerate parametric amplifier could be realized for a microwave mode and an ensemble of NV-centers (or other spins) by parametrically driving the spin ensemble. The parametric driving was sinusoidal at the sum of the spin and cavities frequencies. Here we consider whether the performance of the amplifier can be improved by using a more complex drive. Employing numerical optimization, we find that the optimal driving is primarily a sum of harmonics of the sum frequency. The optimal drive, which is essentially a square wave, ramps up the amplification rate by about $40\%$, while limiting the drive to four harmonics improves the amplification by about $22\%$. 

\end{abstract}

\maketitle

\section{Introduction}

Hybrid microwave--spin platforms combine the low-loss storage and routing capabilities of superconducting resonators with the long coherence times and large collective coupling available in solid-state spin ensembles. Nitrogen-vacancy (NV) centers in diamond are a particularly attractive medium because they provide optically addressable spin states, can be engineered in large ensembles and integrated with circuit-QED architectures and microwave photonics~\cite{Doherty2013,Amsuss2011,Kubo2010, Xiang2013Hybrid}.

In this study, we consider a parametric amplification mechanism discovered in~\cite{OvsiannikovArxiv2601} in which a spin ensemble is parametrically driven, realizing the non-degerate parametric amplifier/squeezer for the ensemble and a microwave cavity. In the weak-excitation (large-spin) regime, the ensemble dynamics can be mapped to a harmonic oscillator via the Holstein--Primakoff approximation~\cite{HolsteinPrimakoff1940, OvsiannikovArxiv2601}, while the fundamental light-matter interaction is  the Jaynes--Cummings model and its collective generalization (Tavis--Cummings)~\cite{JaynesCummings1963,TavisCummings1968}. A time-dependent drive modulates the ensemble transition frequency, effectively realizing a driven cavity-spin Hamiltonian in which resonant modulation can activate two-mode-squeezing-like interactions and thus enable amplification or squeezing of selected field quadratures~\cite{Clerk2010,OvsiannikovArxiv2601}. More broadly, this mechanism places the present device in the context of cavity-based quantum-limited parametric amplification and squeezing, where nondegenerate amplification, input-output relations, and noise constraints play a central role \cite{Caves1982Amplifier,Gardiner1985InputOutput,Bergeal2010JRM}.

Because the performance of a quantum-limited amplifier is ultimately constrained by dissipation, thermal noise, and experimental bounds on the available modulation bandwidth and amplitude, here we explore to what extent the use of a drive containing multiple frequencies can improve the performance of the amplifier. More generally, this problem belongs to the broader framework of quantum control, in which time-dependent driving is engineered to maximize a chosen figure of merit under realistic physical constraints~\cite{Koch2022,Zhang2017,Glaser2015,Petersen2010,Brif2010,dAlessandro2007}. In relatively unconstrained settings, high-fidelity protocols can often be found efficiently by standard gradient-based searches, suggesting a comparatively favorable control landscape~\cite{Brif2010,Rabitz2004}. At the same time, many practically relevant quantum-control tasks involve additional restrictions, and related optimization problems also arise in hybrid quantum--classical settings and variational quantum algorithms~\cite{Magann2021,Ge2022}. This has motivated the use of both quasi-Newton searches such as Broyden-Fletcher-Goldfarb-Shanno (BFGS) algorithm and more specialized optimal-control methods, including Krotov-type schemes, (GRadient Ascent Pulse Engineering), CRAB (Chopped RAndom-Basis), and comparative benchmarking frameworks for control algorithms~\cite{NocedalWright2006,Khaneja2005,Machnes2011,Caneva2011}. For especially complicated control landscapes, where conventional local searches may become inefficient, reinforcement-learning, machine-learning, and Bayesian approaches have been ~\cite{Bukov2018,Mukherjee2020}. Most recently polynomial optimization has been applied to the optimization of quantum control, a method which is able to determine global optima even in complex landscapes densely populated by local extrema~\cite{Bondar2026QCPOP}. We note that because constrained quantum-control problems can
become strongly nonconvex and densely populated by local extrema, when using local optimization methods it can be important to identify efficient control
parametrizations when the search is carried out with local optimization methods~\cite{Bondar2026QCPOP}.

Because the controls considered here are periodic, the problem is also naturally connected to Floquet analysis of driven quantum systems \cite{Shirley1965Floquet}. In addition, the emergence of nearly bang-bang optimal waveforms in bounded-control problems can be understood within the Pontryagin maximum principle and its quantum-control formulations \cite{Boscain2021PMP}.

In the present hybrid cavity-spin platform, the above general quantum control perspective translates into a concrete question: whether the previously considered sinusoidal modulation is already close to optimal, or whether a bounded multi-harmonic waveform can provide a noticeable improvement in amplification and squeezing.

This paper is organized as follows. In Sec.~\ref{sec:system}, we introduce the hybrid cavity-spin system, derive the covariance-matrix equations of motion, and define the figures of merit used to quantify amplification and squeezing. In Sec.~\ref{sec:results}, we present the numerical optimization results, including the properties of the optimal control as well as the amplification and squeezing regimes obtained with both optimal and smoothed driving fields. Finally, in Sec.~\ref{sec:conclusions}, we summarize the main results and discuss their implications for hybrid microwave--spin parametric amplifiers.

\section{System and Method}\label{sec:system}

Let us begin with the mathematical description of  a single-mode microwave oscillator that interacts with a very large number of nitrogen-vacancy centers, which we consider as two-level systems. We write the Lindblad equation for this system in the following form~\cite{Jacobs14, OvsiannikovArxiv2601}:
\begin{align}\label{Lindb_eq}
    \dfrac{d\rho }{dt}& = i[\rho, \tilde{H}] + \frac{\gamma}{2}(1 + n_{T}) \left( 2 a \rho a^\dagger -a^\dagger a \rho - \rho a^\dagger a \right) \notag \\
    & + \frac{\gamma}{2} n_{T} \left( 2 a^\dagger \rho a - a a^\dagger \rho - \rho a a^\dagger \right) \notag\\
    &+ \frac{\kappa}{2}\left( 2 b \rho b^\dagger -b^\dagger b \rho - \rho b^\dagger b \right) ,
\end{align}
where $\tilde{H}$ is the system Hamiltonian divided by $\hbar$, $\gamma$ is the cavity damping rate,  $\kappa$ is the inhomogeneous linewidth of the NV centers~\cite{Fahey23}, $a$ is the photon annihilation operator corresponding to the cavity mode, and $b$ is the annihilation operator for the oscillator that describes the spin ensemble. The operator $b$ is obtained by a limiting transition in the Holstein--Primakoff transformation \cite{HolsteinPrimakoff1940} from spin operators to bosonic operators for a large number of spins in the system. The Hamiltonian~$\tilde{H}$ in Eq.~(\ref{Lindb_eq}) can be expressed as
\begin{equation}
    \tilde{H}= \omega_{\ms{c}} a^\dagger a + \left[\omega_{\ms{s}} + \Lambda  f(t)\right] b^\dagger b + g (a + a^\dagger)(b + b^\dagger),  
    \label{Ham}
\end{equation}
where $\omega_{\ms{c}}$ is the cavity mode frequency, $\omega_{\ms{s}}$ is the frequency of the NV centers in the absence of modulation, $\Lambda$ is the amplitude of the modulation, $g$ is the amplitude of the cavity-spin coupling,  $f(t)$ is a function determining the explicit form of the time-dependent modulation, normalized accordingly, $f(t) \in [-1, 1]$. Finally, $n_T$ is the number of thermal photons present in the resonator at ambient temperature $T$ at the initial moment in time, which is given by the Bose--Einstein distribution function, 
\begin{align}
    n_{T} = \frac{1}{\exp\left[\hbar\omega_c/(k_{\mathrm{B}}T) \right] - 1}, 
\end{align} 
where $k_{\mathrm{B}}$ is the Boltzmann constant. 

Next, let us define the quadrature operators for the cavity mode, $X_a = (a+a^\dagger)/2$, $Y_a = (a-a^\dagger)/(2i)$, and for the NV centers, $X_b = (b+b^\dagger)/2$, $Y_b = (b-b^\dagger)/(2i)$. These allow us to determine the covariance matrix, $ C =  \langle \mathbf{v}\mathbf{v}^{\ms{T}} \rangle - \langle \mathbf{v}\rangle \langle\mathbf{v}^{\ms{T}} \rangle$ 
with $\mathbf{v}^{\ms{T}} \equiv (X_a, Y_a, X_b, Y_b)$, which we analyze in detail below.

Now, using Eq.~(\ref{Lindb_eq}), we arrive at the system of equations that describes the evolution of the covariance matrix of the quadratures:
\begin{equation}\label{cov_eq}
    \frac{\mathrm d C}{\mathrm d t} = A(t) C + C A^{\ms{T}}(t) + G,
\end{equation}
where 
\begin{equation}
    A(t) = 
     \begin{pmatrix}
        - \dfrac{\gamma}{2} & \omega_{\ms{c}} & 0 & 0 \\
        - \omega_{\ms{c}} & -\dfrac{\gamma}{2} & -2 g & 0 \\
        0  & 0  & - \dfrac{\kappa}{2} & \omega_s + \Lambda f(t)  \\
        -2g & 0 & -\omega_s - \Lambda f(t) & - \dfrac{\kappa}{2} 
    \end{pmatrix}
\end{equation}
and
\begin{equation}\label{eq:matr_def}
    G = \frac{1}{4}
     \begin{pmatrix}
       \gamma(2 n_T + 1) & 0 & 0 & 0 \\
        0 & \gamma(2 n_T + 1) & 0 & 0 \\
        0  & 0  & \kappa & 0  \\
        0 & 0 & 0 & \kappa
    \end{pmatrix}. 
\end{equation}

To examine the magnitude of amplification (squeezing) of the first quadrature $\langle \Delta X_a^2 \rangle$, we introduce the quantities $S_{\ms{amp}}$ and $S_{\ms{sqz}}$, measured in dB, in the same way as in Ref~\cite{OvsiannikovArxiv2601},
\begin{eqnarray}
       &&\mathcal{S}\left(V\right) = 10 \log_{10} \left(\! \sqrt{V/V_0} \right) = 10 \log_{10} \left( 2 \sqrt{V} \right) , 
    \\
       &&\mathcal{S}_{\ms{amp}} = \mathcal{S}(V_{\ms{max}}),
       \quad
       \mathcal{S}_{\ms{sqz}} = -\min \left[ 0, \mathcal{S}(V_{\ms{min}}) \right] , 
\end{eqnarray}
where we set the zero of the logarithmic scale to the standard deviations of the quadrature variances for a coherent state, which are all equal to $\sqrt{V_0} = 1/2$.

\section{Results}\label{sec:results}

\subsection{Properties of the optimal control}

With a complete description of the system in terms of Eqs.~\eqref{Lindb_eq}--\eqref{eq:matr_def}, we can now solve the problem of maximizing $\langle \Delta X_a^2 \rangle$ at a fixed point in time $t_f$ by selecting a control function $f(t) \in [-1, 1]$. Optimal control is implemented by means of an external magnetic drive, the shape of which we will select either for maximum signal amplification or for maximum squeezing. To this end, we represent $f(t)$ as a piecewise constant function on a uniform time grid with 40 points per period~$t_p=2\pi/(\omega_c + \omega_s)$, while each value of the control function on this grid is taken as a parameter for optimization. Hence, for the typical timeframe of evolution $t=10^4 t_p$ we have $4\times 10^5$ parameters for optimization. The latter is realized with the Limited-memory Broyden-Fletcher-Goldfarb-Shanno (L-BFGS) method ~\cite{Nocedal1980UpdatingQM, Liu1989OnTL}. 

\begin{figure}
\includegraphics[width=1\columnwidth]{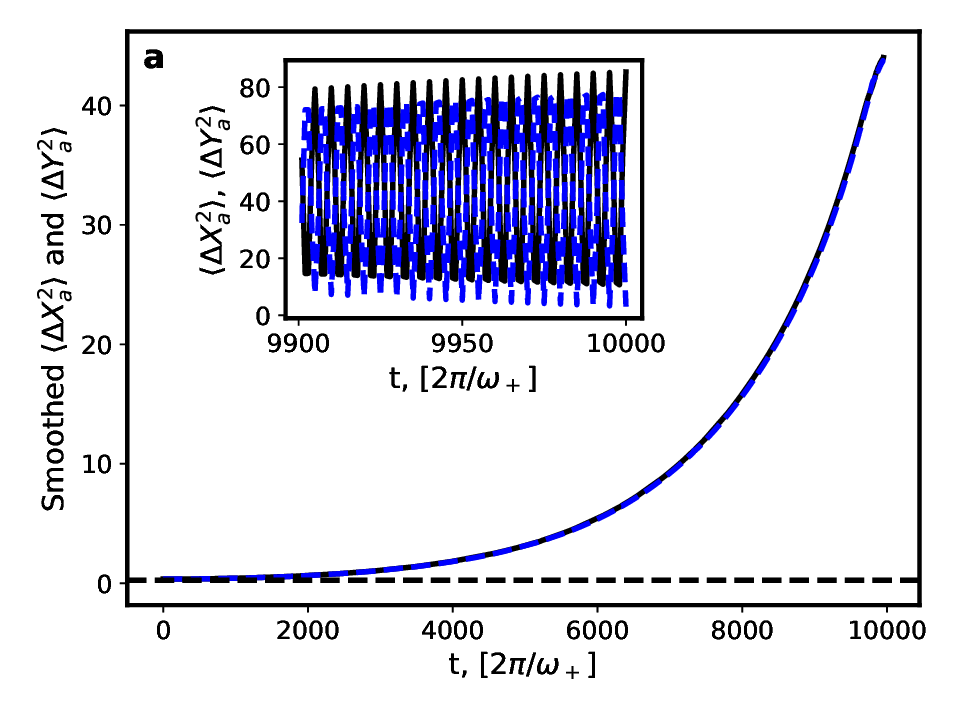}\\
\includegraphics[width=1.055\columnwidth]{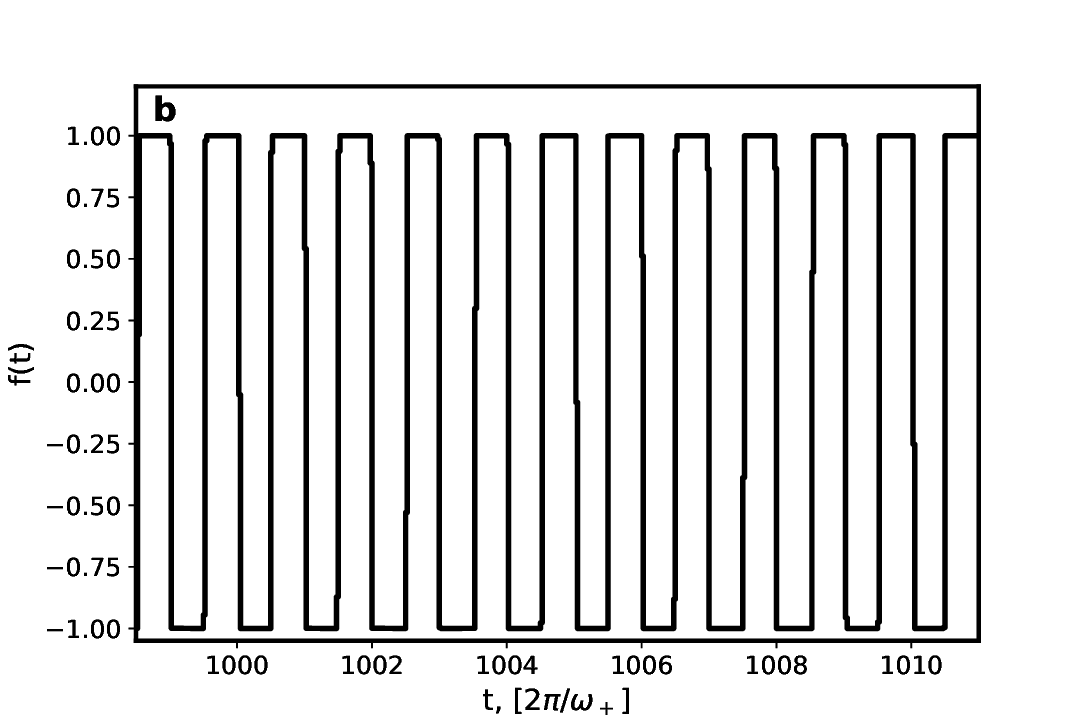}\\
\includegraphics[width=1\columnwidth]{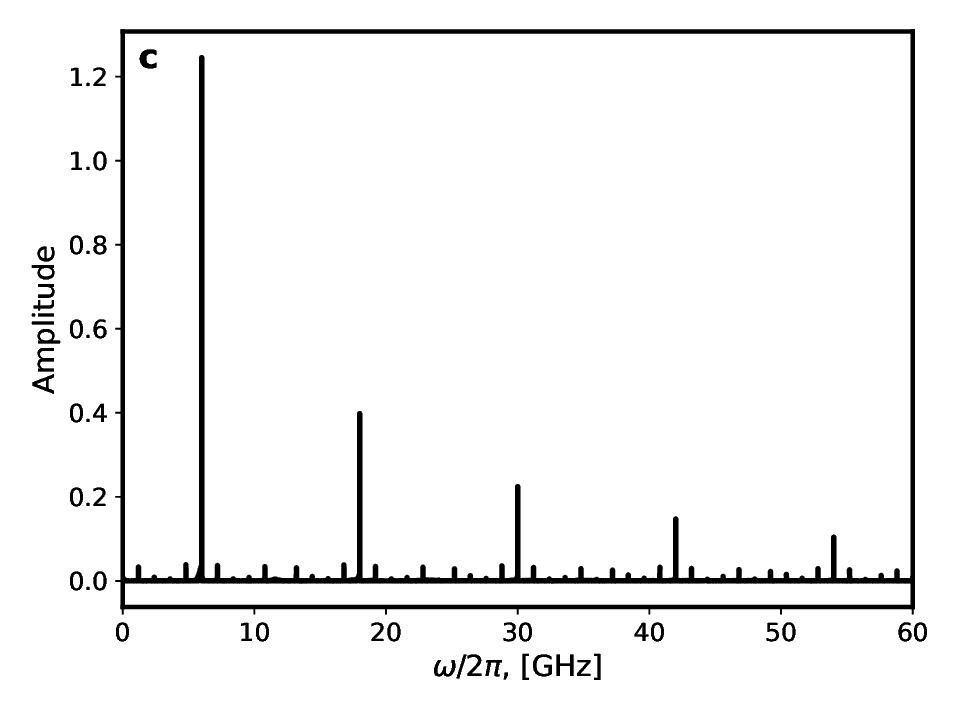} 
    \caption{(a) Time evolution of the smoothed quadratures dispersion $\langle \Delta X_a^2 \rangle$ (black line) and $\langle \Delta Y_a^2 \rangle$ (blue dashed line) maximized by selecting a piecewise-constant function $f(t)$ from Eq.~\eqref{eq:opt_control} with parameters $\omega_c = 2\pi \times 2.4 $~GHz,  $\omega_s = 2\pi \times 3.6 $ GHz, $g = 2 \pi \times 3.5$ MHz, $\gamma = \kappa = 2 \pi \times 200$ kHz, and $T = 10 $ mK.
    Inset: time evolution of the quadratures dispersion $\langle \Delta X_a^2 \rangle$ (black line) and $\langle \Delta Y_a^2 \rangle$ (blue dashed line) 
    over the last 100 periods. (b) Structure of the obtained optimal control function in the selected time interval.
    (c) Spectrum of the obtained optimal control function.}
    \label{fig:glob_opt}
\end{figure}

In Fig.~\ref{fig:glob_opt}, we show the main results of the numerical optimization for the system with the following set of parameters: $\omega_c = 2\pi \times 2.4 $ GHz,  $\omega_s = 2\pi \times 3.6 $ GHz, $g = 2 \pi \times 3.5$~MHz, $\gamma = \kappa = 2 \pi \times 200$~kHz (these parameters are kept fixed throughout this study), $\Lambda = 2\pi \times 1.0$ GHz, and $T = 20$ mK. 
Interestingly, it follows from Fig.~\ref{fig:glob_opt}(b) that the bang-bang control is optimal, which corresponds to the control that enters the system linearly \cite{Sussmann1979BangBang}. It is clear that the signal amplification shown in Fig. ~\ref{fig:glob_opt}(a) is much better than the amplification for a conventional sinusoidal drive described in \cite{OvsiannikovArxiv2601} (a direct comparison is shown in Fig.\ref{fig:opt_ampl}). The resulting control function, shown in Fig. \ref{fig:glob_opt} (b), is a square wave with a constant period throughout the entire evolution, except for a short interval at the beginning and end of the evolution. This is apparently due to the fact that, in the final stages of the search for optimal control, the gradients with respect to the optimization parameters become very small, and a change in the control function toward strict periodicity becomes impossible.

Of the most relevance to us are the results shown in Fig.~\ref{fig:glob_opt}(c). These allow us to draw some conclusions about the properties of the optimal control function. For example, we can see that the spectrum of the optimal control function contains only those peaks that correspond to multiples of the frequencies $\omega_c + \omega_ s$ and $|\omega_c - \omega_ s|$. For convenience, we denote them as follows:
 \begin{equation}
     \omega_{\pm} = |\omega_c \pm \omega_s|.
 \end{equation}

 In fact, we intentionally chose the frequencies $\omega_c$ and $\omega_s$ so that $\omega_+$ is an integer multiple of $\omega_-$. In principle, due to measurement errors for frequencies $\omega_+$ and $\omega_-$ in the experiments, we can always guarantee that, at most, we have the ratio $\omega_+ /\omega_-$ forming a set of positive rational numbers, and in our case, the spectrum of the optimal control function will consist of all possible linear combinations of $\omega_+$ and $\omega_-$ with integer coefficients. This property allows us to find the full period of the optimal control function and use the Floquet approach to reduce the number of parameters to be optimized. In our subsequent simulations, we will treat the frequency ratio $\omega_+ /\omega_-$ as an integer. This is convenient because it allows us to use the smallest frequency, $\omega_-$, as the common period for the Floquet simulation. If the ratio were rational, finding a common multiple frequency that encompasses all the frequencies of the optimal control function would be more complicated.

\subsection{Amplification}

\begin{figure}
\includegraphics[width=1\columnwidth]{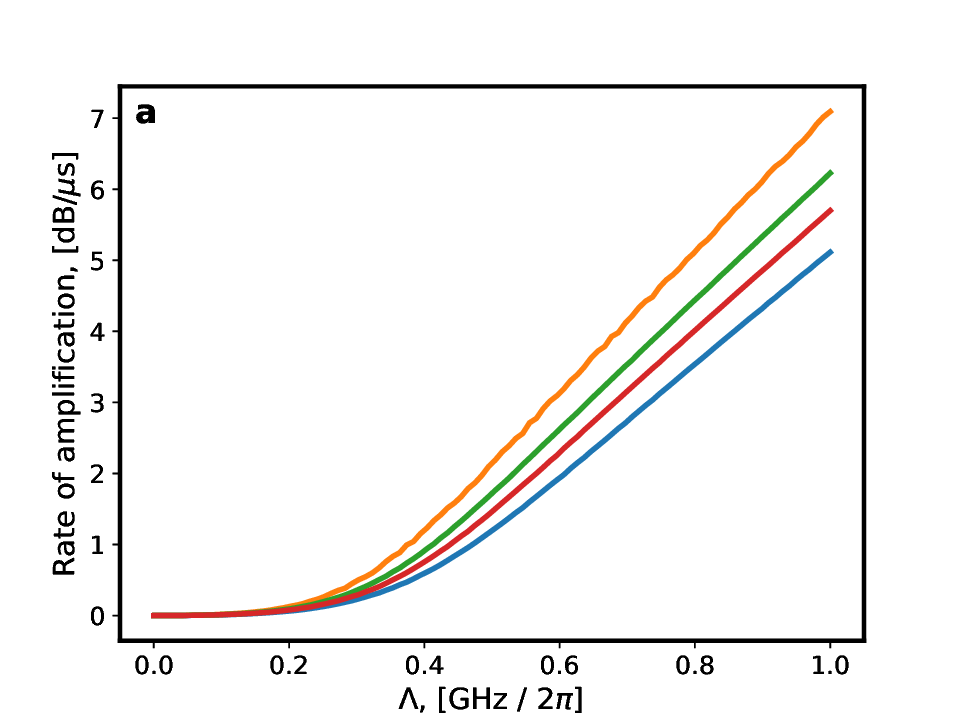}
\includegraphics[width=1\columnwidth]{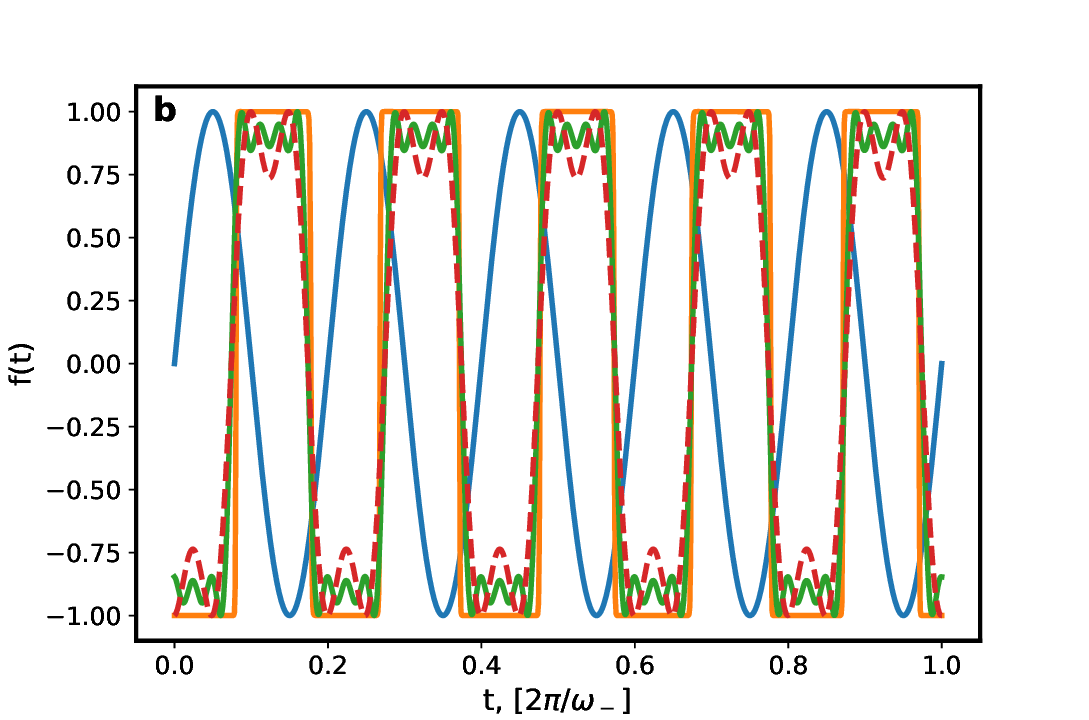} 
    \caption{(a) Amplification rate as a function of the NV modulation amplitude~$\Lambda$ for the sinusoidal (blue line), optimal (orange line), and smoothed optimal driving with $m=2$ (red line) and $m=4$ (green line). Temperature of the cavity mode $T=300$~K, damping rate $\kappa = \gamma = 2 \pi \times 200 ~\mbox{kHz}$. (b) Time modulation of different control types over a single period. The color scheme is the same as in (a). 
    }
    \label{fig:opt_ampl}
\end{figure}

 As we observed in the previous subsection, the optimal control signal shows periodic behavior. This allows us to devise an algorithm requiring optimization of a significantly smaller number of parameters than in its original form. To do this, we will define the control function as a Fourier series in terms of the resonant harmonic $\omega_-$. Since $\omega_+ = m \omega_-$, where $m \in \mathbb{N}$, the harmonics  $\omega_+$ will automatically be taken into account. Next, we can apply the Floquet approach to model the evolution of $\langle \Delta X_a^2 \rangle$ over the period determined by the largest harmonic $\omega_-$. Then it can be represented as follows:
\begin{equation}\label{eq:opt_control}
    f(t) = \tanh \left[\sum_{n = 1}^N \left(a_n \sin(n\omega_- t) + b_n \cos(n\omega_- t)\right)\right],
\end{equation}
where tanh is used to ensure that $f(t) \in [-1, 1]$ without violating the periodicity at the Floquet evolution time we have defined. Next, we can substitute this control into our system of equations for quadratures (\ref{cov_eq}) and perform gradient maximization, for example, using the L-BFGS method. We will search for the control in the form of the first $N = 100$ harmonics in the expansion (\ref{eq:opt_control}). The simulation results compared to sinusoidal driving (blue line) are shown in Fig. \ref{fig:opt_ampl}.
The results of numerical modeling show that the optimal control method (orange line) is very stable in the case of changes in a wide range of system parameters, such as cavity mode temperature, damping coefficients, or parametric driving amplitude. This, in turn, allows us to obtain the  amplification rate for systems with different driving values using essentially the same optimal control and achieve results that are better than in the case of sinusoidal control~\cite{OvsiannikovArxiv2601}. This is apparently due to the fact that the optimal control is very close to the following: 
\begin{equation}
    f_{\ms{amp}}(t) \approx \operatorname{sgn}
    \left[\sin(\omega_+ t + \phi)\right],
\end{equation}
where $\phi$ is a phase that depends weakly on the system parameters (or does not depend on them at all).

However, it is known that bang-bang control is very difficult to achieve under experimental conditions. From the experimental point of view, the implementation of bang-bang control can be challenging due to need to switch the control field on timescales much shorter than the characteristic times of the system's intrinsic dynamics. In practice, standard control electronics often cannot provide sufficiently rapid changes of the applied field, making the realization of a regime close to ideal bang-bang technically demanding. An additional difficulty arises from the need for extremely precise timing synchronization: the switching instants must be matched to the phase and dynamics of the system, since even small timing errors can significantly distort the desired evolution ~\cite{Alonso2016BangBang}. Thus, we consider filtering the original optimal control and examine its main harmonics. By decomposing $f(t)$ into Fourier series and keeping two or four sinusoidal fields, we find that both the resulting drives achieve a higher amplification than that for pure sinusoidal driving. 

The corresponding function for the filtered control, after smoothing, for the chosen set of parameters can be written as the sum of $m$ modes,
\begin{equation}\label{eq:smtd}
    f_{smtd}^{(m)}(t) = \sum_{i=1}^m a_i \cos(q_i\omega_- t + \phi_i).
\end{equation}
In particular, for $m=2$ (red lines in Fig.\ref{fig:opt_ampl}) and for the above-mentioned system parameters, we obtain $q_1=5$, $q_2=15$, $a_1 = 1.07$, $a_2 = 0.34$, $\phi_1 = 2.39$, and $\phi_2 = -2.26$. If we keep more harmonics, e.g., $m=4$, of the initial optimal signal (green lines in Fig.~\ref{fig:opt_ampl}), we obtain a gain rate that is even closer to the values obtained for the optimal control. At the same time, the set of parameters corresponding to this control function is $q_i=\{5,15,25,35\}$, $a_i = \{1.14,0.36,0.1,0.11\}$, and $\phi_i = \{2.39,-2.26,-0.61,1.03\}$.

\begin{figure}
    \centering
\includegraphics[width=1\columnwidth]{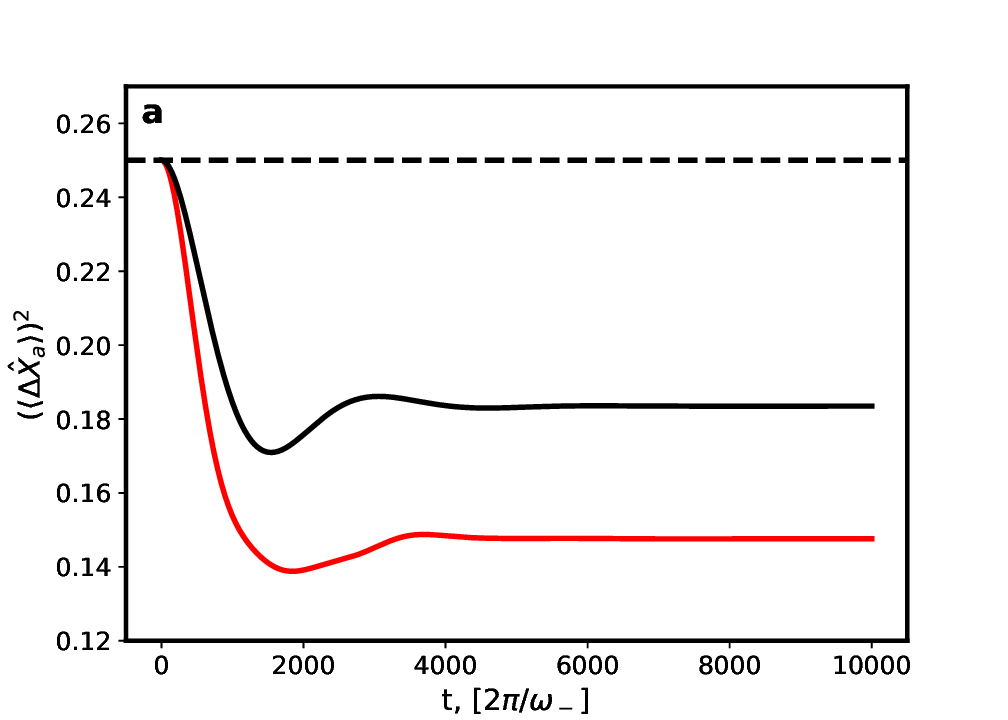}
\includegraphics[width=1\columnwidth]{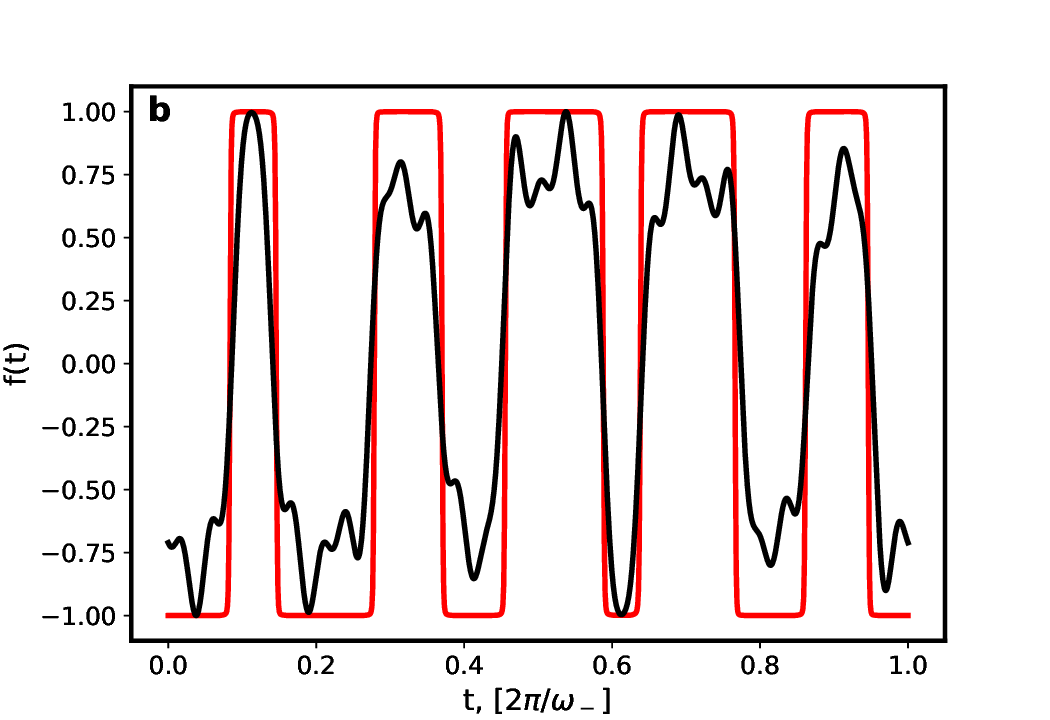} 
    \caption{(a) Time evolution of $\langle \Delta X_a^2 \rangle$ for the optimal control (red line) and smoothed optimal control with $m=6$ (black line). Other parameters are $\Lambda = 2\pi \times 0.5~\mbox{GHz}$, $\kappa = \gamma = 2 \pi \times 200 ~\mbox{kHz}$, and $T=1~\mbox{mK}$. (b) Detailed structure of the optimal control function (red line) and smoothed optimal control (black line) over a single period $2\pi/\omega_{-}$. 
    }
    \label{fig:opt_sq}
\end{figure}

\subsection{Squeezing}

Our method of constructing optimal control allows us to achieve squeezing for a separate quadrature (in our case, $\langle \Delta X_a^2 \rangle$). 
While this is an improvement over sinusoidal driving \cite{OvsiannikovArxiv2601}, the value of squeezing that can be obtained is very modest. For parameters that can be implemented in a physical system (albeit with a rather large driving amplitude, $\Lambda$), the squeezing is about $1.2$ dB, as shown in Fig.~\ref{fig:opt_sq}(a) (red line). At the same time, from Fig.~\ref{fig:opt_sq}(b), we see that the optimal control retains the bang-bang structure (red line). The main contrast with the protocol for amplification is that here modulation in time is less symmetric and, accordingly, less stable with respect to changes in system parameters. 

Again, as in the case of amplification, let us apply the smoothed optimal control. The results of such modeling for signal squeezing are also shown in Fig. \ref{fig:opt_sq} (black lines), where we keep the first six harmonics of the optimal signal. As we can see from Fig.~\ref{fig:opt_sq}(a), the resulting squeezing is lower than that obtained with bang-bang control; in particular, it is approximately 0.65~dB, which is almost half of that generated by the optimal control. The parameters for our smoothed control as defined by Eq.(\ref{eq:smtd}) are: $q_i=\{1, 5, 9, 11, 17, 27\}$, $a_i = \{0.32, 0.81, 0.24, 0.23, 0.15, 0.11\}$, and $\phi_i = \{2.53, 2.72, -0.25, 1.43, 0.50, 2.75\}$.

\section{Conclusions}\label{sec:conclusions}

We have studied a hybrid microwave-spin platform in which a single cavity mode is coupled to the collective bosonic mode of a large NV-center ensemble whose frequency is modulated. We employed optimal control to design bounded driving waveforms for the modulation that maximize either quadrature amplification or squeezing. 

Our simulations show that the optimal control enables an improvement in amplifier gain over sinusoidal driving~\cite{OvsiannikovArxiv2601} of about $40 \%$, with the numerically optimal solution approaching a bang-bang profile. The obtained spectra of the optimal driving are dominated by harmonics at integer combinations of $\omega_{+}=|\omega_{c}+\omega_{s}|$ and $\omega_{-}=|\omega_{c}-\omega_{s}|$, thereby motivating a compact Fourier-parameterized control ansatz. 

Since ideal bang-bang switching is challenging to implement under experimental conditions, we also demonstrate that filtering the optimal waveform to a small number of leading harmonics yields smooth, experimentally more feasible control that preserves a significant fraction of the achievable amplification. At the same time, for the squeezing regime, the attainable reduction of $\langle\Delta X_a^2\rangle$ is rather moderate (on the order of 1.2~dB), while the optimal form of driving modulation becomes more sensitive to parameter variations.

\acknowledgments

D.I.B. was supported by DEVCOM Army Research Office (ARO) (grant W911NF-23-1-0288; program manager Dr.~James Joseph). R.O. and A.G.S. acknowledge support by the National Research Foundation of Ukraine, project No.~2023.03/0073. The views and conclusions contained in this document are those of the authors and should not be interpreted as representing the official policies, either expressed or implied, of ARO or the U.S. Government. The U.S. Government is authorized to reproduce and distribute reprints for Government purposes notwithstanding any copyright notation herein.

\bibliography{refs.bib}

\end{document}